\documentclass[11pt,a4paper]{article}
\newcommand{\s}{\sigma}
\newcommand{\la}{\lambda}

\begin{document}

\title{\bf {Quintom model with O($N$) symmetry}}
\author{\normalsize{M. R. Setare$^{1}$\thanks{%
E-mail: rezakord@ipm.ir}  \, and \,E.~N.~Saridakis $^{2}$\thanks{%
E-mail: msaridak@phys.uoa.gr} }\\
\newline
\\
{\normalsize \it $^1$ Department of Science, Payame Noor
University, Bijar, Iran}
\\
{\normalsize \it $^2$ Department of Physics, University of Athens,
GR-15771 Athens, Greece}
\\
}
\date{\small{}}

\maketitle
\begin{abstract}

We investigate the quintom model of dark energy in the generalized
case where the corresponding canonical and phantom fields possess
O($N$) symmetries. Assuming exponential potentials we find that
this O$(N)$ quintom paradigm exhibits novel properties comparing
to the simple canonical and phantom scenarios. In particular, we
find that the universe cannot result in a quintessence-type
solution with $w>-1$, even in the cases where the phantom field
seems to be irrelevant. On the contrary, there are always
late-time attractors which correspond to accelerating universes
with $w<-1$ and with a recent crossing of the phantom divide, and
for a very large area of the parameter space they are the only
ones. This is in contrast with the previous simple-quintom
results, where an accelerating universe is a possible late-time
stable solution but it is not guaranteed.

 \end{abstract}

\newpage

\section{Introduction}\

Many cosmological observations, such as SNe Ia \cite{1}, WMAP
\cite{2}, SDSS \cite{3}, Chandra X-ray observatory \cite{4}, etc.,
reveal that our universe is undergoing an accelerating expansion.
In addition, they suggest that it is spatially flat, and consists
of about $70\%$ of dark energy with negative pressure, of $30\%$
of dust matter (cold dark matter plus baryons), and of an
negligible amount of radiation. Although the nature and origin of
dark energy could perhaps understood by a fundamental underlying
theory unknown up to now, physicists can still propose some
paradigms to describe it. The most obvious theoretical candidate
for dark energy is the cosmological constant \cite{5,Weinberg89,6}
which has the equation of state $w =-1$. However, it leads to the
two known difficulties \cite{7}, namely the ``fine-tuning''
problem (why is the current vacuum energy density so small), and
the ``cosmic coincidence'' one (why are the densities of vacuum
energy and dark matter nearly equal today since they  scale very
differently during the expansion history).

There have been many efforts to resolve these problems
\cite{Weinberg89,tries}, but none could offer a robust and
undoubted solution. This fact led many theoretical physicists to
construct alternative frameworks, such is the dynamical dark
energy scenario, by assuming that the vacuum energy is cancelled
to exactly zero by some unknown mechanism, and introducing a dark
energy component with a dynamically variable equation of state.
The dynamical dark energy proposal is often realized by some
scalar-field mechanism which suggests that the energy form with
negative pressure is provided by a scalar field evolving downwards
a proper potential. A large class of scalar-field dark energy
models have been studied in the literature, including quintessence
\cite{quintessence}, K-essence \cite{kessence}, tachyon
\cite{tachyon}, phantom \cite{10}, ghost condensate \cite{ghost2},
holographic dark energy \cite{holoext}, bulk holographic dark
energy \cite{bulkhol} and many others.

A primary scalar field candidate for dark energy was the
quintessence scenario \cite{quintessence}, which consists of a
fluid with equation-of-state parameter lying in the range $-1< w<
{-1/3}$. On the other hand, for the phantom model \cite{10} of
dark energy, which consists of a scalar field with a negative sign
of the kinetic term in the Lagrangian, one always obtains $w\leq
-1$. Thus, neither the quintessence nor the phantom alone can
fulfill the transition from $w>-1$ to $w<-1$ and vice versa,
although the cosmological observations mildly favor models where
such a transition was indeed realized and in particular with $w$
crossing $-1$ in the near past. As it was indicated in the
literature \cite{11}, the consideration of the combination of
quintessence and phantom in a unified model, leads to the
fulfillment of the aforementioned transition through the $w=-1$
divide. This model, dubbed quintom, can produce a better fit to
the observational data.

The generalization of quintessence and phantom models to fields
with O$(N)$ symmetry have been performed in \cite{Li02} and
\cite{13} respectively. As it was shown, the behavior of the
corresponding dynamical systems, in specific areas of the
parameter space, can be qualitatively different than then
single-field models. In the present work we are interested in
investigating the generalized quintom model with an $O(N)$
symmetry. As a specific potential form we impose the exponential
dependence on the corresponding fields, since exponential
potentials are known to be significant in various cosmological
models \cite{Copeland}. We perform a complete phase-space
stability analysis of the corresponding autonomous system and we
extract its attractor properties. We find that the O$(N)$ quintom
has qualitatively novel properties compared to the  corresponding
simple phantom, quintessence and quintom models. In particular,
cosmological solutions with $w<-1$ are the only attractors for a
very large area of the parameter space, and the crossing through
the $w=-1$ divide has a large probability to be realized.

The plan of the work is as follows: In section \ref{ONquintom} we
construct the quintom model with O($N$) symmetries and in section
\ref{ONstab} we perform its complete stability analysis. In
section \ref{cosmimpl}  we discuss the cosmological implications
of our results, and finally section \ref{conclusions} is devoted
to conclusions.

\section{O($N$) quintom}
\label{ONquintom}

We consider a flat Robertson-Walker metric:
\begin{equation}\label{metric}
ds^{2}=dt^{2}-a^{2}(t)d\textbf{x}^2.
\end{equation}
The Lagrangian density for a quintom model with
 O($N$) symmetries is:
\begin{equation}
L=\frac{1}{2}g^{\mu\nu}\left[(\partial_{\mu}\Phi^{\alpha})(\partial_{\nu}\Phi^{\alpha})-(\partial_{\mu}\s^{\beta})(\partial_{\nu}\s^{\beta})\right]-
V_\Phi(|\Phi^{\alpha}|)-V_\s(|\s^{\beta}|),
\end{equation}
where $\Phi^{\alpha}$ is the component of the canonical field,
with $\alpha=1,2,\cdots,N_\Phi$, and  $\s^{\beta}$ is the
component of the phantom field, with $\beta=1,2,\cdots,N_\s$. Note
that in the general case, the dimensionality of the multiplets of
the two fields, is not the same. Fortunately, these
dimensionalities do not appear in the final form of the equations.

 In order to impose the O($N$)
symmetries, following \cite{Li02}, we write:
\begin{eqnarray}\label{imag}
\Phi^{1}&=&R_\Phi(t)\cos\varphi_{\Phi_1}(t)\nonumber\\
\Phi^{2}&=&R_\Phi(t)\sin\varphi_{\Phi_1}(t)\cos\varphi_{\Phi_2}(t)\nonumber\\
\Phi^{3}&=&R_\Phi(t)\sin\varphi_{\Phi_1}(t)\sin\varphi_{\Phi_2}(t)\cos\varphi_{\Phi_3}(t)\\
&\cdots\cdots&\nonumber\\
\Phi^{N_\Phi-1}&=&R_\Phi(t)\sin\varphi_{\Phi_1}(t)\cdots\sin\varphi_{\Phi_{N_\Phi-2}}(t)\cos\varphi_{\Phi_{N_\Phi-1}}(t)\nonumber\\
\Phi^{N_\Phi}&=&R_\Phi(t)\sin\varphi_{\Phi_1}(t)\cdots\sin\varphi_{\Phi_{N_\Phi-2}}(t)\sin\varphi_{\Phi_{N_\Phi-1}}(t)\nonumber,
\end{eqnarray}
and similarly:
\begin{eqnarray}\label{imag}
\s^{1}&=&R_\s(t)\cos\varphi_{\s_1}(t)\nonumber\\
\s^{2}&=&R_\s(t)\sin\varphi_{\s_1}(t)\cos\varphi_{\s_2}(t)\nonumber\\
\s^{3}&=&R_\s(t)\sin\varphi_{\s_1}(t)\sin\varphi_{\s_2}(t)\cos\varphi_{\s_3}(t)\\
&\cdots\cdots&\nonumber\\
\s^{N_\s-1}&=&R_\s(t)\sin\varphi_{\s_1}(t)\cdots\sin\varphi_{\s_{N_\s-2}}(t)\cos\varphi_{\s_{N_\s-1}}(t)\nonumber\\
\s^{N_\s}&=&R_\s(t)\sin\varphi_{\s_1}(t)\cdots\sin\varphi_{\s_{N_\s-2}}(t)\sin\varphi_{\s_{N_\s-1}}(t)\nonumber.
\end{eqnarray}
Thus, we have explicitly used the properties
$|\Phi^{\alpha}|=R_\Phi$ and  $|\s^{\beta}|=R_\s$. Furthermore, we
assume that the potentials $V_\Phi(|\Phi^{\alpha}|)$ and
$V_\s(|\s^{\beta}|)$ depend only on $R_\Phi$ and $R_\s$
respectively.

The action for the universe is as usual:
\begin{equation}\label{action}
S=\int d^{4}x\sqrt{-g}\left(-\frac{1}{16\pi
G}R_{s}-p_{\gamma}+L\right),
\end{equation}
where $g$ is the determinant of the metric tensor $ g_{\mu\nu}$,
$R_{s}$ is the Ricci scalar and $G$ is the Newton's constant (in
the following we will instead use $\kappa^2\equiv8\pi G$).
$p_{\gamma}$ is the pressure of the barotropic fluid which
constitutes the matter content of the universe, with equation of
state $p_{\gamma}=(\gamma-1)\rho_{\gamma}$, with the constant
$\gamma$ in the interval $0<\gamma<2$.

The Einstein equations for the angles $\varphi_{\Phi_\alpha}$ and
$\varphi_{\s_\beta}$ can be easily derived through the
corresponding variation of the action. However, they are
irrelevant for the purpose of this work since we are going to use
only the radial equations for the fields, i.e. the equations
determining the evolution of $R_\Phi(t)$ and $R_\s(t)$, plus the
Friedmann equations. Hence, we have:
\begin{equation}
H^{2}=\frac{\kappa^2}{3}\left[\rho_{\gamma}+\rho_{\Phi}+\rho_{\s}\right],
\label{Fr1}
\end{equation}
\begin{equation}
\left(\frac{\ddot{a}}{a}\right)=-\frac{\kappa^2}{3}\left[\left(\frac{3\gamma}{2}-1\right)\rho_{\gamma}+2
p_{\Phi}+2 p_{\s}+V_\Phi(R_\Phi)+V_\s(R_\s)\right], \label{Fr2}
\end{equation}
\begin{equation}
\ddot{R}_\Phi+3H\dot{R}_\Phi-\frac{\Omega_\Phi^{2}}{a^{6}R_\Phi^{3}}-\frac{\partial
V_\Phi(R_\Phi)}{\partial R_\Phi}=0,
\label{canonical}
\end{equation}
\begin{equation}
\ddot{R}_\s+3H\dot{R}_\s-\frac{\Omega_\s^{2}}{a^{6}R_\s^{3}}-\frac{\partial
V_\s(R_\s)}{\partial R_\s}=0. \label{phantom}
\end{equation}
In equations (\ref{Fr1}),(\ref{Fr2}), the energy density and
pressure of the canonical and the phantom fields, are given by:
\begin{eqnarray}
\rho_{\Phi}=\frac{1}{2}(\dot{R}_\Phi^{2}+\frac{\Omega_\Phi^{2}}{a^{6}R_\Phi^{2}})+V_\Phi(R_\Phi)\nonumber\\
p_{\Phi}=\frac{1}{2}(\dot{R}_\Phi^{2}+\frac{\Omega_\Phi^{2}}{a^{6}R_\Phi^{2}})-V_\Phi(R_\Phi)\label{enercanon}
\end{eqnarray}
and
\begin{eqnarray}
\rho_{\s}=-\frac{1}{2}(\dot{R}_\s^{2}+\frac{\Omega_\s^{2}}{a^{6}R_\s^{2}})+V_\s(R_\s)\nonumber\\
p_{\s}=-\frac{1}{2}(\dot{R}_\s^{2}+\frac{\Omega_\s^{2}}{a^{6}R_\s^{2}})-V_\s(R_\s).\label{enerphantom}
\end{eqnarray}
In addition, the effect of the ``angular component'' of the system
is embedded in the radial equations
(\ref{canonical}),(\ref{phantom}) as an effective term containing
the constants $\Omega_\Phi$ and $\Omega_\s$, which are determined
by the values of the first integrals of motion \cite{Li02}.
Finally, $H$ is Hubble parameter.

The equation of state for the O$(N)$ quintom is:
\begin{equation}
w =\frac{p_\Phi+p_\s}{\rho_\Phi+\rho_\s}=
\frac{\dot{R}_\Phi^{2}-\dot{R}_\s^{2}+\frac{1}{a^6}
\left(\frac{\Omega_\Phi^{2}}{R_\Phi^{2}}-\frac{\Omega_\s^{2}}{R_\s^{2}}\right)-2[V_\Phi(R_\Phi)+V_\s(R_\s)]}
{\dot{R}_\Phi^{2}-\dot{R}_\s^{2}+\frac{1}{a^6}
\left(\frac{\Omega_\Phi^{2}}{R_\Phi^{2}}-\frac{\Omega_\s^{2}}{R_\s^{2}}\right)+2[V_\Phi(R_\Phi)+V_\s(R_\s)]}\label{eqstate}.
\end{equation}
Thus, the constructed quintom model could produce a value $w<-1$
if
\begin{equation}
\left|\dot{R}_\Phi^{2}-\dot{R}_\s^{2}+\frac{1}{a^6}
\left(\frac{\Omega_\Phi^{2}}{R_\Phi^{2}}-\frac{\Omega_\s^{2}}{R_\s^{2}}\right)\right|<2\left|V_\Phi(R_\Phi)+V_\s(R_\s)\right|.
\end{equation}

\section{Stability analysis of the O$(N)$ quintom}
\label{ONstab}

We are interested in  investigating the attractor properties of
the O$(N)$ quintom model, imposing exponential potentials, since
they are known to be relevant in various cosmological models
\cite{Copeland}. In particular we consider:
\begin{eqnarray}
V_\Phi(R_\Phi)=V_{\Phi_0}\exp(-\lambda\kappa R_\Phi)\nonumber\\
V_\s(R_\s)=V_{\s_0}\exp(-\lambda\kappa R_\s).
\end{eqnarray}
Using this specific potential ansatz, the radial equations of
motion (\ref{canonical}),(\ref{phantom}) become:
\begin{equation}
\ddot{R}_\Phi+3H\dot{R}_\Phi-\frac{\Omega_\Phi^{2}}{a^{6}R_\Phi^{3}}-\la
\kappa\, V_\Phi(R_\Phi)=0, \label{canonical2}
\end{equation}
\begin{equation}
\ddot{R}_\s+3H\dot{R}_\s-\frac{\Omega_\s^{2}}{a^{6}R_\s^{3}}-\la
\kappa\, V_\s(R_\s)=0. \label{phantom2}
\end{equation}
Furthermore, using the definitions for the energy densities and
pressures (\ref{enercanon}),(\ref{enerphantom}), the Friedmann
equations (\ref{Fr1}),(\ref{Fr2}) can be re-written as:
\begin{equation}
\label{sys1}
\dot{H}=-\frac{\kappa^2}{2}\left(\rho_\gamma+p_\gamma+\dot{R}_\Phi^2
+ \frac{\Omega_\Phi^2}{a^6R_\Phi^2}-\dot{R}_\s^2 -
\frac{\Omega_\s^2}{a^6R_\s^2}\right)
\end{equation}
\begin{equation}\label{sys2}
H^2=\frac{\kappa^2}{3}\left[\rho_\gamma+\frac{1}{2}\left(\dot{R_\Phi}^2+\frac{\Omega_\Phi^2}{a^6R_\Phi^2}\right)+V_\Phi(R_\Phi)-
\frac{1}{2}\left(\dot{R_\s}^2+\frac{\Omega_\s^2}{a^6R_\s^2}\right)+V_\s(R_\s)\right].
\end{equation}
Finally, the equations close by considering the evolution of the
barotropic (matter) density:
\begin{equation}\label{sys3}
\dot{\rho_\gamma}=-3H(\rho_\gamma+p_\gamma).
\end{equation}

In order to perform the stability analysis of the O$(N)$ quintom
model, we have to transform the dynamical system
(\ref{canonical2})-(\ref{sys2}) into an autonomous form
\cite{Hilbert}. This will be achieved by introducing the auxiliary
variables:
\begin{eqnarray}
x_\Phi=\frac{\kappa}{\sqrt{6}H}\dot{R}_\Phi &,&\ \  x_\s=\frac{\kappa}{\sqrt{6}H}\dot{R}_\s\nonumber\\
y_\Phi=\frac{\kappa\sqrt{V_\Phi(R_\Phi)}}{\sqrt{3}H} &,&\ \  y_\s=\frac{\kappa\sqrt{V_\s(R_\s)}}{\sqrt{3}H}\nonumber\\
z_\Phi=\frac{\kappa}{\sqrt{6}H}\frac{\Omega_\Phi}{a^3R_\Phi} &,&\ \  z_\s=\frac{\kappa}{\sqrt{6}H}\frac{\Omega_\s}{a^3R_\s}\nonumber\\
\xi_\Phi=\frac{1}{\kappa R_\Phi} &,&\ \  \xi_\s=\frac{1}{\kappa
R_\s}\label{auxilliary},
\end{eqnarray}
together with $M=\log a$.

Using these variables, we result in the following autonomous
system:
\begin{eqnarray}\label{auto}
&&\frac{dx_\Phi}{dM}=\frac{3}{2}x_\Phi T_1-
3x_\Phi+\sqrt{6}\,z_\Phi^2\xi_\Phi+\sqrt{\frac{3}{2}}\la\, y_\Phi^2\nonumber\\
&&\frac{dx_\s}{dM}=\frac{3}{2}x_\s T_1-
3x_\s+\sqrt{6}\,z_\s^2\xi_\s-\sqrt{\frac{3}{2}}\la\, y_\s^2\nonumber\\
&&\frac{dy_\Phi}{dM}=\frac{3}{2}y_\Phi T_1-
\sqrt{\frac{3}{2}}\la\,x_\Phi y_\Phi \nonumber\\
&&\frac{dy_\s}{dM}=\frac{3}{2}y_\s T_1-
\sqrt{\frac{3}{2}}\la\, x_\s y_\s \nonumber\\
&&\frac{dz_\Phi}{dM}=\frac{3}{2}z_\Phi T_1-
3z_\Phi-\sqrt{6}\,x_\Phi z_\Phi\xi_\Phi\nonumber\\
&&\frac{dz_\s}{dM}=\frac{3}{2}z_\s T_1-
3z_\s-\sqrt{6}\,x_\s z_\s\xi_\s\nonumber\\
&&\frac{d\xi_\Phi}{dM}=-\sqrt{6}\,\xi_\Phi^2x_\Phi\nonumber\\
&&\frac{d\xi_\s}{dM}=-\sqrt{6}\,\xi_\s^2x_\s\label{autonomous},
\end{eqnarray}
where
$T_1=\gamma\left(1-x_\Phi^2-y_\Phi^2-z_\Phi^2+x_\s^2-y_\s^2+z_\s^2\right)+2(x_\Phi^2+z_\Phi^2-x_\s^2-z_\s^2)$.
Note also that the Friedmann equation (\ref{sys2}) leads to the
constraint equation:
\begin{equation}\label{constraint}
x_\Phi^2-x_\s^2+y_\Phi^2+y_\s^2+z_\Phi^2-z_\s^2+\frac{\kappa^2\rho_{\gamma}}{3H^2}=1.
\end{equation}
Finally, in terms of the auxiliary variables, the equation of
state for the quintom (\ref{eqstate}) becomes:
\begin{equation}\label{eqstate2}
 w=\frac{x_\Phi^2-x_\s^2-y_\Phi^2-y_\s^2+z_\Phi^2-z_\s^2}{x_\Phi^2-x_\s^2+y_\Phi^2+y_\s^2+z_\Phi^2-z_\s^2}.
\end{equation}

The critical points $(x_{\Phi c},x_{\s c},y_{\Phi c},y_{\s
c},z_{\Phi c},z_{\s c},\xi_{\Phi c},\xi_{\s c})$ of the autonomous
system (\ref{autonomous}) are obtained by setting the left hand
sides of the equations to zero. The real and physically meaningful
of them are presented in table \ref{crit}.
\begin{table*}[h]
\begin{center}
\begin{tabular}{|c|c|c|c|c|c|c|c|c|}
\hline
 Cr. Point& $x_{\Phi c}$ & $x_{\s c}$ & $y_{\Phi c}$ & $y_{\s c}$ & $z_{\Phi c}$ & $z_{\s c}$ &  $\xi_{\Phi c}$ & $\xi_{\s c}$   \\
\hline \hline
 A&  $+\sqrt{1+x_\s^2}$ & $x_\s$ & 0 & 0 &   0 & 0  &   0 & 0\\
\hline
 B&  $-\sqrt{1+x_\s^2}$ & $x_\s$ & 0 & 0 &   0 & 0  &   0 & 0\\
\hline
C& $\frac{\sqrt{6}}{\la}$ & $+\frac{\sqrt{6-\la^2}}{\la}$ & 0 & 0 &   0 & 0  &   0 & 0\\
\hline
D& $\frac{\sqrt{6}}{\la}$ & $-\frac{\sqrt{6-\la^2}}{\la}$ & 0 & 0 &   0 & 0  &   0 & 0\\
\hline
 E& $\frac{\la}{\sqrt{6}}$ & 0 & $+\sqrt{1-\frac{\la^2}{6}}$ & 0 &   0 & 0  &   0 & 0\\
\hline
 F& $\frac{\la}{\sqrt{6}}$ & 0 & $-\sqrt{1-\frac{\la^2}{6}}$ & 0 &   0 & 0  &   0 & 0\\
\hline
 G& $\sqrt{\frac{3}{2}}\frac{\gamma}{\la}$ & 0 & $+\sqrt{\frac{3}{2}}\frac{\sqrt{\gamma(2-\gamma)}}{\la}$ & 0 &   0 & 0  &   0 & 0\\
\hline
 H& $\sqrt{\frac{3}{2}}\frac{\gamma}{\la}$ & 0 & $-\sqrt{\frac{3}{2}}\frac{\sqrt{\gamma(2-\gamma)}}{\la}$ & 0 &   0 & 0  &   0 & 0\\
\hline
 I& $+\frac{\sqrt{\la^2+6}}{\la}$  & $\frac{\sqrt{6}}{\la}$  & 0 & 0 &   0 & 0  &   0 & 0\\
\hline
 J& $-\frac{\sqrt{\la^2+6}}{\la}$  & $\frac{\sqrt{6}}{\la}$  & 0 & 0 &   0 & 0  &   0 & 0\\
\hline
 K& 0 & $-\frac{\la}{\sqrt{6}}$  & 0 & $+\frac{\sqrt{\la^2+6}}{\sqrt{6}}$  &   0 & 0  &   0 & 0\\
\hline
L& 0 & $-\frac{\la}{\sqrt{6}}$   & 0 & $-\frac{\sqrt{\la^2+6}}{\sqrt{6}}$  &   0 & 0  &   0 & 0\\
\hline
\end{tabular}
\end{center}
\caption[crit]{\label{crit} The real and physically meaningful
critical points of the autonomous system (\ref{autonomous}).}
\end{table*}

In order to determine the stability properties of these critical
points, we proceed as follows: We expand the auxiliary variables
around the critical points as
\begin{eqnarray}
x_\Phi&=&x_{\Phi c}+u_\Phi\nonumber\\
x_\s&=&x_{\s c}+u_\s\nonumber\\
y_\Phi&=&y_{\Phi c}+v_\Phi\nonumber\\
y_\s&=&y_{\s c}+v_\s\nonumber\\
z_\Phi&=&z_{\Phi c}+w_\Phi\nonumber\\
z_\s&=&z_{\s c}+w_\s\nonumber\\
\xi_\Phi&=&\xi_{\Phi c}+\chi_\Phi\nonumber\\
\xi_\s&=&\xi_{\s c}+\chi_\s
 \label{expansion}.
\end{eqnarray}
In expressions (\ref{expansion}),
$u_\Phi,u_\s,v_\Phi,v_\s,w_\Phi,w_\s,\chi_\Phi,\chi_\s$ are just
the perturbations of the variables near the critical points and we
consider them forming a column vector denoted as $\textbf{U}$.
Inserting these expansions into the autonomous system
(\ref{autonomous}), we can obtain the equations for the
perturbations up to first order as:
\begin{eqnarray}
\label{perturbation}
\textbf{U}'={\bf{Q}}\cdot \textbf{U},
\end{eqnarray}
where the prime denotes differentiation with respect to $M$.

 For
the critical points $(x_{\Phi c},x_{\s c},y_{\Phi c},y_{\s
c},z_{\Phi c},z_{\s c},\xi_{\Phi c},\xi_{\s c})$, the coefficients
of the perturbation equations form a $8\times8$ matrix ${\bf
{Q}}$. Thus, for each critical point of table \ref{crit}, the
eigenvalues of ${\bf {Q}}$ determine the type and stability of
this specific critical point. The explicit form of the matrix
${\bf {Q}}$ and its eigenvalues for the critical points of table
\ref{crit}, i.e for the autonomous system (\ref{autonomous}), are
given in the Appendix. In table \ref{stability} we present the
results of the stability analysis. In addition, for each critical
point we calculate the values of $w$ (given by relation
(\ref{eqstate2})), and of $\frac{\kappa^2\rho_\gamma}{3H^2}$
(given by the constraint (\ref{constraint})). Thus,
$\frac{\kappa^2\rho_\gamma}{3H^2}=0$ means that the universe is
dominated completely by the quintom fields.  In the next section
we discuss the cosmological implications of the obtained results.
\begin{table*}[h]
\begin{center}
\begin{tabular}{|c|c|c|c|c|}
\hline
 Cr. Point& Existence & Stability & $w$ & $\frac{\kappa^2\rho_\gamma}{3H^2}$   \\
\hline \hline
 A &   $\forall$ $\gamma$ and  $\forall$ $\la$ &
Stable node for {\tiny{$\la x_{\s }>\sqrt{6}$}} &1&0 \\ & &
Unstable node for {\tiny{$-\sqrt{6-\la^2}<\la x_{\s
}<\sqrt{6-\la^2}$}}
  &  &  \\ & & Saddle point
for {\tiny{$\la x_{\s }<-\sqrt{6-\la^2}$}} or
{\tiny{$\sqrt{6-\la^2}<\la x_{\s }<\sqrt{6}$}}
  &  & \\
\hline
 B&  $\forall$ $\gamma$ and  $\forall$ $\la$ & Unstable node & 1 & 0 \\
\hline
C& $0<\la^2<6$ & Unstable node & 1 & 0 \\
\hline
D& $0<\la^2<6$  & Unstable node& 1 & 0 \\
\hline
 E& $\la^2<6$ & Saddle point & $-1+\la^2/3$&0 \\
\hline
  F& $\la^2<6$ & Saddle point&$-1+\la^2/3$&0 \\
\hline
 G&  $\forall$ $\gamma$ and  $\la\neq0$ & Saddle point & $\gamma-1$ & $1-3\gamma/\la^2$ \\
\hline
 H&  $\forall$ $\gamma$ and $\la\neq0$ & Saddle point & $\gamma-1$ & $1-3\gamma/\la^2$ \\
\hline
 I&  $\forall$ $\gamma$ and  $\la\neq0$  & Saddle point  & 1 & 0 \\
\hline
 J&  $\forall$ $\gamma$ and   $\la\neq0$  & Unstable node & 1 & 0 \\
\hline
 K& $\forall$ $\gamma$ and    $\forall$ $\la$ & Stable node & $-1-\la^2/3$ & 0\\
\hline
 L& $\forall$ $\gamma$ and    $\forall$ $\la$  & Stable node  & $-1-\la^2/3$ & 0 \\
\hline
\end{tabular}
\end{center}
\caption[stability]{\label{stability} The properties of the
critical points of the autonomous system (\ref{autonomous}).}
\end{table*}

\section{Cosmological implications}
\label{cosmimpl}

The critical points K and L are stable nodes for every $\gamma$
and every $\la$ and thus they constitute late-time attractors.
Since $\frac{\kappa^2\rho_\gamma}{3H^2}=0$ they correspond to
quintom-dominated universes, with the quintom equation-of-state
parameter being  $w=-1-\frac{\la^2}{3}$. The fact that $x_{\Phi
c}=y_{\Phi c}=0$ implies that the phantom component of the quintom
plays the main role in this solution. Thus, this solution
corresponds to the phantom solution of the literature, with the
expected $w<-1$. However, the more complex dynamics of the quintom
model makes the system qualitative different from the simple
phantom one, even in this particular solution  where the canonical
field seems to have a trivial contribution. Indeed, in the simple
phantom case \cite{13} a stable node exists only when $\la^2<6$,
which in turn leads to a lower bound for $w$. As we see this is
not true in the O($N$) quintom paradigm, where the stable node is
independent of the parameter values. Thus, although in the
previous studies an accelerating universe was not always a
late-time attractor, in the present analysis we show that such a
cosmological era is always a stable equilibrium solution. In
addition, the possible values of $w$ are not bounded.

The critical points A and B are a novel property of the present
quintom model. First of all, since for every $x_{\s}$ there is
always a $x_{\Phi}$ which lead to zero left hand sides of the
autonomous system (\ref{autonomous}) (given the nullification of
the other variables), A and B corresponds to loci of critical
points. They correspond to quintom-dominated solutions, with
$w=1$. Note that in this case both the canonical and the phantom
field are relevant for the cosmological evolution. The critical
point B is unstable. However, in the case $\la x_{\s }>\sqrt{6}$,
the critical point A is a stable node. This behavior is different
form both the simple canonical \cite{Copeland} and the simple
phantom \cite{13} fields.

The critical points  C, D and I correspond to quintom-dominated
solutions, and especially to solutions where the constraint
(\ref{constraint}) is dominated only by the kinetic energy of the
canonical and the phantom fields. As expected, these
``kinetic''-dominated solutions are unstable and are only expected
to be relevant at early times. Furthermore, the critical point J
corresponds to a quintom-dominated unstable solution.

The critical points G and H correspond to saddle points of the
dynamical system at hand, for every $\gamma$ and every $\la\neq0$.
Note that although for a specific combination of these parameters
($\gamma>\frac{2}{9}$ and $\la^2>\frac{24\gamma^2}{9\gamma-2}$)
two of the eigenvalues form a complex conjugate pair with negative
real part, the corresponding critical points are not stable
spirals but saddle points, due to the presence of a positive
eigenvalue. This is a crucial difference comparing to the simple
canonical field \cite{Copeland} and reveals that the more complex
quintom dynamics changes qualitatively the behavior of the system,
even if the phantom field does not seem to play an important role
in this case ($x_{\s c}=y_{\s c}=0$). The value of $w$ for the
critical points G and H is $\gamma-1$, thus for the allowed
$\gamma$ values it corresponds to $-1<w<1$. In other words, this
solution corresponds to the quintessence universe and the fact
that is not a stable point, i.e. a late-time attractor, means that
in the model at hand the universe cannot result in a
quintessence-type solution with $w>-1$, even in the cases where
the phantom field seems to be irrelevant. Finally, we mention that
for these points
$\frac{\kappa^2\rho_\gamma}{3H^2}=1-\frac{3\gamma}{\la^2}$. Thus,
although G and H exist for every $\gamma$ and every $\la\neq0$,
the corresponding solution has a physical meaning only when
$\la^2>3\gamma$. This fact was also mentioned in \cite{Copeland}
for the simple canonical model.

The critical points E and F, are saddle points. They correspond to
quintom-dominated universes, with  $w=-1+\frac{\la^2}{3}$. The
fact that $x_{\s c}=y_{\s c}=0$ means that the phantom field is
not relevant for this solution, and the cosmological evolution is
driven by the canonical field. Thus, this solution also
corresponds to the quintessence universe. However, while this
solution is a stable node, i.e. a late-time attractor, for the
simple quintessence model in the case $\la^2<3\gamma$
\cite{Copeland}, in the present quintom model it is always a
saddle point. Therefore, the quintessence-type solution cannot be
an equilibrium point for the cosmological evolution of the
universe. Once again we see that the more complex O($N$) quintom
dynamics implies qualitatively new behavior for the cosmological
evolution comparing to the previous simple models.

From the aforementioned investigation we conclude that the O($N$)
quintom has a variety of critical points. It is interesting to see
that for not very steep potentials and/or a small $x_{\s}$, the
attractor A disappears. Thus, the attractors K and L are the only
ones for a very large area of the two-dimensional parameter space
($\gamma,\la$). And since the initial value of $w$ can be easily
above $-1$ we conclude that the transition through the $w=-1$
divide has a large probability to be realized, as the system is
attracted be the attractors K and L.

\section{Conclusions}
\label{conclusions}

In the present work we investigated the quintom paradigm of dark
energy in the generalized case where the corresponding canonical
and phantom fields possess O($N$) symmetries. Our analysis reveals
that this O$(N)$ quintom model exhibits novel properties comparing
to the simple canonical and phantom scenarios. In particular, we
find that the universe cannot result in a quintessence-type
solution with $w>-1$, even in the cases where the phantom field
seems to be irrelevant. On the contrary, there are always
late-time attracting equilibrium points which correspond to
quintom-dominated, accelerating universes with $w<-1$, and for a
very large area of the parameter space they are the only ones.
This is in contrast with the previous simple-quintom results,
where an accelerating universe is a possible late-time stable
solution but it is not guaranteed. In addition, the more complex
quintom dynamics, refutes the lower bound of $w$ obtained
previously in the simple phantom paradigm, making its possible
values unlimited. Finally, the transition form above to below of
the phantom divide $w=-1$, has a large probability to be realized.
These features make the present O($N$) quintom model consistent
with cosmological observations.

\section*{Appendix: stability of the critical points}

For the critical points $(x_{\Phi c},x_{\s c},y_{\Phi c},y_{\s
c},z_{\Phi c},z_{\s c},\xi_{\Phi c},\xi_{\s c})$, the coefficients
of the perturbation equations form a $8\times8$ matrix ${\bf
{Q}}$, which for the autonomous system (\ref{autonomous}) reads:

\begin{eqnarray}
{\bf {Q}}_{11}&=&3\left[-1+\frac{T_2}{2}+x_{\Phi c}^2(2-\gamma)\right]\nonumber\\
{\bf {Q}}_{12}&=&3\,x_{\Phi c}\,x_{\s c}\,(\gamma-2)\nonumber\\
{\bf {Q}}_{13}&=&y_{\Phi c}(\sqrt{6}\la-3\gamma\, x_{\Phi c})\nonumber\\
{\bf {Q}}_{14}&=&-3\,\gamma\, x_{\Phi c}\,y_{\s c}\nonumber\\
{\bf {Q}}_{21}&=&3 \,x_{\Phi c}\,x_{\s c}\,(2-\gamma)\nonumber\\
{\bf {Q}}_{22}&=&3\left[-1+\frac{T_2}{2}+x_{\s c}^2(\gamma-2)\right]\nonumber\\
{\bf {Q}}_{23}&=&-3\,\gamma\, x_{\s c}\,y_{\Phi c} \nonumber\\
{\bf {Q}}_{24}&=&y_{\s c}(-\sqrt{6}\la-3\,\gamma \,x_{\s c})\nonumber\\
{\bf {Q}}_{31}&=&\left[-\sqrt{\frac{3}{2}}\la+3\,x_{\Phi
c}(2-\gamma)\right]y_{\Phi c}\nonumber\\
{\bf {Q}}_{32}&=&3\,x_{\s c}\,y_{\Phi c}\,(\gamma-2)\nonumber\\
{\bf {Q}}_{33}&=&-\sqrt{\frac{3}{2}}\la\, x_{\Phi c}+\frac{3T_2}{2}-3\,\gamma \,y_{\Phi c}^2\nonumber\\
{\bf {Q}}_{34}&=&-3\,\gamma\, y_{\Phi c}\, y_{\s c}\nonumber\\
{\bf {Q}}_{41}&=&3\,x_{\Phi c}\,y_{\s c}\,(2-\gamma)\nonumber\\
{\bf {Q}}_{42}&=&\left[-\sqrt{\frac{3}{2}}\la+3\,x_{\s c}\,(\gamma-2)\right]y_{\s c}\nonumber\\
{\bf {Q}}_{43}&=&-3\,\gamma\, y_{\Phi c} \,y_{\s c}\nonumber\\
{\bf {Q}}_{44}&=&-\sqrt{\frac{3}{2}}\la \,x_{\s c}+\frac{3T_2}{2}-3\,\gamma\, y_{\s c}^2\nonumber\\
{\bf {Q}}_{55}&=&3\left(-1+\frac{T_2}{2}\right)\nonumber\\
{\bf {Q}}_{66}&=&3\left(-1+\frac{T_2}{2}\right)\nonumber
 \label{Qmatrix},
\end{eqnarray}
where $T_2=\gamma(1-x_{\Phi c}^2+x_{\s c}^2-y_{\Phi c}^2-y_{\s
c}^2)+2(x_{\Phi c}^2-x_{\s c}^2)$. All the other components of
${\bf {Q}}$ are zero for the physically interested critical points
of table \ref{crit}.

The eigenvalues $\nu_i$ ($i=1,\cdots,8$) of ${\bf {Q}}$ for the
specific critical points of table \ref{crit}, i.e for the
autonomous system (\ref{autonomous}), are presented in table
\ref{eigen}, where
$T_3=\frac{1}{\la^2}\sqrt{\la^2(2-\gamma)(24\gamma^2+2\la^2-9\gamma\la^2)}$.
Thus, by determining the sign of the real parts of these
eigenvalues, we can classify the corresponding critical point
\cite{Hilbert}.
\begin{table*}[h]
\begin{center}
\begin{tabular}{|c|c|c|c|c|c|c|c|c|}
\hline
 Cr. Point & $\nu_1$ & $\nu_2$ & $\nu_3$ & $\nu_4$ & $\nu_5$ & $\nu_6$ & $\nu_7$& $\nu_8$  \\
\hline \hline
 A & $3(2-\gamma)$ & $3-\sqrt{\frac{3}{2}}\la\, x_{\s}$  & $3-\sqrt{\frac{3}{2}}\la \sqrt{1+x^2_{\s}}$   & 0 & 0 & 0 & 0 & 0\\
\hline
B & $3(2-\gamma)$ & $3-\sqrt{\frac{3}{2}}\la\, x_{\s}$  & $3+\sqrt{\frac{3}{2}}\la \sqrt{1+x^2_{\s}}$   & 0 & 0 & 0 & 0 & 0\\
\hline
 C &  $6-3\gamma$ & $3-3\sqrt{1-\frac{\la^2}{6}}$  & 0 & 0 & 0 & 0 & 0 & 0\\
\hline
D&  $6-3\gamma$ & $3+3\sqrt{1-\frac{\la^2}{6}}$  & 0 & 0 & 0 & 0 & 0 & 0\\
\hline
 E &  $\frac{\la^2}{2}$ & $\frac{\la^2}{2}-3$  & $\frac{\la^2}{2}-3$  &  $\frac{\la^2}{2}-3$ &  $\frac{\la^2}{2}-3$ & $\frac{\la^2}{2}-3\gamma$  & 0 & 0 \\
\hline
F &  $\frac{\la^2}{2}$ & $\frac{\la^2}{2}-3$  & $\frac{\la^2}{2}-3$  &  $\frac{\la^2}{2}-3$ &  $\frac{\la^2}{2}-3$ & $\frac{\la^2}{2}-3\gamma$  & 0 & 0 \\
\hline
G & $\frac{3}{2}\gamma$ &
 $\frac{3}{4}\left(\gamma-2-T_3\right)$ & $\frac{3}{4}\left(\gamma-2+T_3\right)$ & $\frac{3}{2}(\gamma-2)$ &
  $\frac{3}{2}(\gamma-2)$  &  $\frac{3}{2}(\gamma-2)$   & 0 & 0 \\
\hline
H & $\frac{3}{2}\gamma$ &
 $\frac{3}{4}\left(\gamma-2-T_3\right)$ & $\frac{3}{4}\left(\gamma-2+T_3\right)$ & $\frac{3}{2}(\gamma-2)$ &
  $\frac{3}{2}(\gamma-2)$  &  $\frac{3}{2}(\gamma-2)$   & 0 & 0 \\
\hline
 I &  $6-3\gamma$ & $3-\sqrt{\frac{3}{2}}\sqrt{6+\la^2}$  & 0   &  0  & 0   & 0    & 0  & 0  \\
\hline
 J&  $6-3\gamma$ & $3+\sqrt{\frac{3}{2}}\sqrt{6+\la^2}$  & 0   &  0  & 0   & 0    & 0  & 0  \\
\hline
 K &  $-\frac{\la^2}{2}$ & $-\frac{\la^2}{2}-3$  & $-\frac{\la^2}{2}-3$  & $-\frac{\la^2}{2}-3$ & $-\frac{\la^2}{2}-3$  & $-\la^2-3\gamma$ & 0 & 0\\
\hline
 L &  $-\frac{\la^2}{2}$ & $-\frac{\la^2}{2}-3$  & $-\frac{\la^2}{2}-3$  & $-\frac{\la^2}{2}-3$ & $-\frac{\la^2}{2}-3$  & $-\la^2-3\gamma$ & 0 &
 0\\
\hline
\end{tabular}
\end{center}
\caption[stability]{\label{eigen} The eigenvalues of the matrix
${\bf {Q}}$ of the perturbation equations of the autonomous system
(\ref{autonomous}).}
\end{table*}


\begin{thebibliography}{99}

\bibitem{1}
A. G.  Riess {\it{et al.}} [Supernova Search Team Collaboration],
Astrophys. J. {\bf 607}, 665 (2004) [arXiv:astro-ph/0402512];
 R. A. Knop
{\it{et al.}}, [Supernova Cosmology Project Collaboration],
Astrophys. J. {\bf 598}, 102 (2003) [arXiv:astro-ph/0309368]; S.
Perlmutter {\it{et al.}} [Supernova Cosmology Project
Collaboration], Astrophys. J. {\bf 517}, 565 (1999)
[arXiv:astro-ph/9812133].

\bibitem{2}
C. L. Bennett {\it{et al.}}, Astrophys. J. Suppl. {\bf 148}, 1
(2003) [arXiv:astro-ph/0302207]; D. N. Spergel {\it{et al.}},
Astrophys. J. Suppl. {\bf 148}, 175 (2003)
[arXiv:astro-ph/0302209].

\bibitem{3}
M. Tegmark {\it{et al.}} [SDSS Collaboration], Phys. Rev. D {\bf
69}, 103501 (2004) [arXiv:astro-ph/0310723]; U. Seljak {\it{et
al.}}, Phys. Rev. D {\bf 71}, 103515 (2005) [astro-ph/0407372]; J.
K. Adelman-McCarthy {\it{et al.}} [SDSS Collaboration],
[arXiv:astro-ph/0507711].

\bibitem{4}
S. W. Allen, {\it{et al.}}, Mon. Not. Roy. Astron. Soc. {\bf 353},
457 (2004) [astro-ph/0405340].

\bibitem{5}A.~Einstein,  {\it {Cosmological considerations in the General Theory of Relativity}}, Sitzungsber
Preuss,
Akad. Wiss. Berlin (Math. Phys.), 142 (1917).

\bibitem{Weinberg89}
 S.~Weinberg, Rev. Mod. Phys. {\bf 61}, 1 (1989).

\bibitem{6}
V.~Sahni and A.~Starobinsky, Int. J. Mod. Phys. D {\bf 9}, 373
(2000) [arXiv:astro-ph/9904398]; P.~J.~Peebles  and B.~Ratra,
 Rev. Mod. Phys.  {\bf 75}, 559 (2003) [arXiv:astro-ph/0207347].

\bibitem{7}P.~J.~Steinhardt,  {\it {Critical Problems in Physics}} (1997), Princeton
University Press.

\bibitem{tries}
B.~Zumino, Nucl. Phys. B {\bf 89}, 535 (1975); Z.~Kakushadze,
Nucl. Phys. B {\bf 589}, 75 (2000); R.~Bousso and J.~Polchinski,
JHEP {\bf 0006}, 006 (2000);  S.~Kachru, M.~Schulz and
E.~Silverstein, Phys. Rev. D {\bf 62}, 045021 (2000);
F.~K.~Diakonos and E.~N.~Saridakis, [arXiv:0708.3143[hep-th]].

\bibitem{quintessence}
B.~Ratra and P.~J.~E.~Peebles, Phys.\ Rev.\ D {\bf 37}, 3406
(1988); C.~Wetterich, Nucl.\ Phys.\ B {\bf 302}, 668 (1988);
R.~R.~Caldwell, R.~Dave and P.~J.~Steinhardt, Phys.\ Rev.\ Lett.\
{\bf 80}, 1582 (1998) [arXiv:astro-ph/9708069]; A.~R.~Liddle and
R.~J.~Scherrer, Phys.\ Rev.\ D {\bf 59}, 023509 (1999)
[arXiv:astro-ph/9809272]; I.~Zlatev, L.~M.~Wang and
P.~J.~Steinhardt, Phys.\ Rev.\ Lett.\ {\bf 82}, 896 (1999)
[arXiv:astro-ph/9807002].

\bibitem{kessence}
C.~Armendariz-Picon, V.~F.~Mukhanov and P.~J.~Steinhardt, Phys.\
Rev.\ Lett.\  {\bf 85}, 4438 (2000) [arXiv:astro-ph/0004134].

\bibitem{tachyon}
  A.~Sen,
  JHEP {\bf 0207}, 065 (2002)
  [arXiv:hep-th/0203265];
M. R. Setare, Phys. Lett. B {\bf 653}, 116, (2007).

\bibitem{10} R. R. Caldwell, Phys. Lett. B {\bf 545}, 23 (2002);
S. Nojiri and S. D. Odintsov, Phys. Lett. B {\bf 562}, 147 (2003)
[arXiv:hep-th/0303117];  Y.H. Wei and Y. Tian, Class. Quant. Grav.
{\bf 21}, 5347 (2004) [arXiv:gr-qc/0405038]; V.K. Onemli and R. P.
Woodard, Phys.\ Rev.\ D {\bf 70}, 107301 (2004)
[arXiv:gr-qc/0406098]; S. Capozziello, S. Nojiri, S. D. Odintsov,
Phys. Lett. B {\bf 632}, 597 (2006); M. R. Setare, Eur. Phys. J. C
{\bf 50}, 991 (2007).

\bibitem{ghost2}
F.~Piazza and S.~Tsujikawa, JCAP {\bf 0407}, 004 (2004)
[arXiv:hep-th/0405054].

\bibitem{holoext}
M.~Li, Phys.\ Lett.\ B {\bf 603}, 1 (2004); K.~Enqvist and
M.~S.~Sloth, Phys.\ Rev.\ Lett.\  {\bf 93}, 221302 (2004)
[arXiv:hep-th/0406019]; E.~Elizalde, S.~Nojiri, S.~D.~Odintsov and
P.~Wang, Phys.\ Rev.\ D {\bf 71}, 103504 (2005)
[arXiv:hep-th/0502082]; M. R. Setare, Phys. Lett.{\bf B} 642, 1,
(2006); M. R. Setare, J. Zhang and X. Zhang, JCAP {\bf0703}, 007
(2007) [arXiv:gr-qc/0611084].

\bibitem{bulkhol}
E.~N.~Saridakis, Phys. Lett. B {\bf 660}, 138 (2008)
[arXiv:0712.2228[hep-th]]; E.~N.~Saridakis, JCAP {\bf{0804}}, 020,
(2008) [arXiv:0712.2672[astro-ph]]; E.~N.~Saridakis, Phys. Lett. B
{\bf 661}, 335 (2008)
 [arXiv:0712.3806[gr-qc]].


\bibitem{11}Z. K. Guo, {\it{et al.}},
Phys. Lett. B {\bf608}, 177 (2005) [arXiv:astro-ph/0410654]; G.-B.
Zhao, J.-Q. Xia, B. Feng and X. Zhang, Int. J. Mod. Phys. D
{\bf16}, 1229 (2007)  [arXiv:astro-ph/0603621]; J.-Q. Xia, B. Feng
and X. Zhang, Mod. Phys. Lett. A {\bf 20}, 2409 (2005); B. Feng,
M. Li, Y.-S. Piao and X. Zhang, Phys. Lett. B {\bf634}, 101
(2006); M. R. Setare, Phys. Lett. B {\bf 641}, 130 (2006); M. R.
Setare, J. Sadeghi, and A. R. Amani, Phys. Lett. B {\bf 660}, 299
(2008); J. Sadeghi, M. R. Setare , A. Banijamali and F. Milani,
Phys. Lett. B {\bf662}, 92 (2008); M. R. Setare and E. N.
Saridakis, [arXiv:0802.2595 [hep-th]]; M. R. Setare and E. N.
Saridakis, [arXiv:0807.3807 [hep-th]].


\bibitem{Li02}
X.~Li, J.~Hao and D.~Liu, Class. Quant. Grav. {\bf{19}}, 6049
(2002) [arXiv:astro-ph/0107171].

\bibitem{13}X. Li and J. Hao, Phys. Rev. D {\bf{69}}, 107303 (2004) [arXiv:hep-th/0303093].

\bibitem{Copeland}
E.~J.~Copeland, A.~R.~Liddle and D.~Wands, Phys.\ Rev.\ D {\bf
57}, 4686 (1998)  [arXiv:gr-qc/9711068].

\bibitem{Hilbert} R. Courant and D. Hilbert, {\it {Methods of
Mathematical Physics}}, Vol. 2, Cambridge University Press.




\end{thebibliography}
\end{document}